\documentclass[doublecol]{epl2}

\usepackage{amsmath,amssymb}
\usepackage{color}
\usepackage{mathrsfs}

\newcommand{\void}[1]{}

\title{Universal patterns in multifrequency-driven dissipative systems}
\shorttitle{Universal patterns in multifrequency-driven dissipative systems}

\author{M. L. Olivera-Atencio \inst{1} \and L. Lamata \inst{2} \and S.
Kohler \inst{3} \and J. Casado-Pascual \inst{1}}
\shortauthor{M. L. Olivera-Atencio \etal}
\institute{ 
\inst{1}{F\'isica Te\'orica, Universidad de Sevilla, Apartado de Correos 1065, 41080 Sevilla, Spain}  
\\                 
\inst{2}{Departamento de F\'isica At\'omica, Molecular y Nuclear, Universidad de Sevilla, 41080 Sevilla, Spain}
\\
\inst{3} {Instituto de Ciencia de Materiales de Madrid, CSIC, Cantoblanco, 28049 Madrid, Spain}
}

\date{\today}

\abstract{The response of dissipative systems to multi-chromatic fields
exhibits generic properties which follow from the discrete
time-translation symmetry of each driving component. We derive these
properties and illustrate them with paradigmatic examples of classical and
quantum dissipative systems. In addition, some computational aspects, in
particular a matrix continued-fraction method, are discussed. Moreover, we
propose possible implementations with quantum optical settings.
}


\begin{document}

\maketitle

\section{Introduction}

The dynamics of strongly driven systems may be rather complex, in
particular when the driving consists of various components with different
frequencies.  A particular and well-studied case is a higher harmonic added
to the driving with a fundamental frequency.  This type of bichromatic
driving can be used to control spatio-temporal symmetries via the relative
phase between the two monochromatic fields \cite{ReimannPRL97, FlachPRL00,
LehmannJCP03, KohlerPR05, KohlerEPJB20}.  It allows one to induce directed
motion by the action of an oscillating field with zero mean, giving rise to
the celebrated ratchet effect \cite{ReimannPR02, HanggiRMP09}.  Moreover,
one may use bichromatic driving for quantum state
preparation~\cite{GomezLeonPRR20, Long21}.

By contrast, there exists considerably less work on driving forces with two
incommensurable frequencies, i.e., frequencies whose proportion is an
irrational number.  An intriguing feature of such drivings is that while
each of its components is time-periodic, the system as a whole lacks
discrete time-translation symmetry.  In spite of this, the response may
have higher symmetry than in the commensurable case \cite{ForsterPRB15b,
KohlerEPJB20}.
The reason for this is that, in the incommensurable case, the long-time
average is equivalent to the average over the relative phases among the
driving components \cite{CasadoPascualPRE15, OliveraAtencioEPJB20}.  In
Ref.~\cite{ForsterPRB15b} the difference between commensurable and
incommensurable drivings has been demonstrated both theoretically and
experimentally for the electron transport through a double quantum dot.

At first sight, the distinction between drivings with commensurable and
incommensurable frequencies seems surprising, since irrational numbers can
be approximated to any degree of accuracy by rational numbers.  This
apparent paradox has motivated several studies of the response of
multi-chromatically driven systems---both dissipative
\cite{CasadoPascualPRE13, CuberoPRL14, CasadoPascualPRE15,
OliveraAtencioEPJB20} and Hamiltonian ones \cite{CuberoPRL18,
CuberoPRE18}---as a function of one driving frequency while keeping the
others constant.

In this perspective, we shed light on how the discrete time translation
symmetries of each periodic driving component provides resonance peaks with
a generic shape.  These results are illustrated with some paradigmatic
examples of classical and quantum dissipative systems.  Moreover, we
demonstrate that the phase-average of the long-time response can be
computed with a matrix continued-fraction method originally developed for
incommensurable frequencies \cite{ForsterPRB15b}.  Possible implementations
in quantum optical systems are also suggested together with an outlook for
further studies.

\section{Some general theoretical results}

We are interested in systems whose dynamical equations depend on time
through $N$ time-periodic functions of the form
\begin{equation}	
	f_j(t)=\epsilon_j \cos\left(\Omega_j t+\varphi_j\right),		
	\label{functions}
\end{equation}
where $j=1,\dots,N$, and  $\epsilon_j$, $\Omega_j$, and $\varphi_j$ denote,
respectively, the amplitude, the angular frequency, and the initial phase
of $f_j(t)$. The state of the system at time $t$ will be denoted as
$\mathscr{S}(t)$. Depending on the case, $\mathscr{S}$ may represent the
values of a finite number of state variables characterizing a classical
deterministic system, the density operator of a quantum system, the one-time
probability density of a classical stochastic system, etc.

We focus on generic properties that do not depend on
the precise nature of the functions $f_j(t)$ and the specific details of
the underlying dynamics. Our only
assumption is that there exists a unique steady state,
$\smash{\mathscr{S}^{\mathrm{st}}(t)}$, to which the system converges in
the long-time limit---an assumption that holds for a wide class of
dissipative systems. In general, the steady state will depend on the
specific values taken by the parameters appearing in the functions
$f_j(t)$. When necessary, this dependence will be made explicit by the
notation $\smash{\mathscr{S}^{\mathrm{st}}(t, \boldsymbol{\epsilon},
\boldsymbol{\Omega}, \boldsymbol{\varphi})}$,
where $\boldsymbol{\epsilon}$, $\boldsymbol{\Omega}$, and
$\boldsymbol{\varphi}$ are $N$-dimensional vectors with components
$\epsilon_j$, $\Omega_j$, and $\varphi_j$, respectively.

Since we are assuming a unique steady state, its time evolution must be
uniquely determined by the dynamical equations.  Hence, the steady state
shares the symmetry properties of the dynamical equations. To be specific, the
set of functions in Eq.~\eqref{functions} is invariant under the
$N+1$ transformations
\begin{eqnarray}
	\mathcal{T}^{(j)} &:& \{t,\boldsymbol{\epsilon},\boldsymbol{\Omega},\boldsymbol{\varphi}\}\mapsto\{t,\boldsymbol{\epsilon}^{(j)},\boldsymbol{\Omega},\boldsymbol{\varphi}+\pi  \boldsymbol{u}^{(j)}\},\label{trans1}\\
	\mathcal{T}&:& \{t,\boldsymbol{\epsilon},\boldsymbol{\Omega},\boldsymbol{\varphi}\}\mapsto \{t+\tau,\boldsymbol{\epsilon},\boldsymbol{\Omega},\boldsymbol{\varphi}-\tau \boldsymbol{\Omega}\},\label{trans2}
\end{eqnarray}
where $\smash{\boldsymbol{\epsilon}^{(j)}}$ is the vector of amplitudes
with the sign of component $j$ inverted, while
$\smash{\boldsymbol{u}^{(j)}}$ is the $j$th canonical basis vector.
More formally, 
$\smash{\epsilon^{(j)}_k=(1-2 \delta_{j,k})\epsilon_k}$ and
$\smash{u_k^{(j)}=\delta_{j,k}}$, with $\delta_{j,k}$ being the
Kronecker delta. Since the only explicit dependence of the dynamical
equations on $t$, $\boldsymbol{\epsilon}$, $\boldsymbol{\Omega}$, and
$\boldsymbol{\varphi}$ comes from the functions $f_j(t)$, the steady state
will also be invariant under these $N+1$ transformations, i.e., 
\begin{align}
	\mathscr{S}^{\mathrm{st}}(t,\boldsymbol{\epsilon},\boldsymbol{\Omega},\boldsymbol{\varphi})&=\mathscr{S}^{\mathrm{st}}(t,\boldsymbol{\epsilon}^{(j)},\boldsymbol{\Omega},\boldsymbol{\varphi}+\pi  \boldsymbol{u}^{(j)})\label{sym1}\\
	&=\mathscr{S}^{\mathrm{st}}(t+\tau,\boldsymbol{\epsilon},\boldsymbol{\Omega},\boldsymbol{\varphi}-\tau \boldsymbol{\Omega})\label{sym2}.
\end{align}

\section{Generic shape of the resonance peaks}

Let $Q=Q(\mathscr{S})$ represent a certain (physical) quantity that depends
on the state of the system. In particular, in the steady state,  the
dependence of $Q$ on $t$, $\boldsymbol{\epsilon}$, $\boldsymbol{\Omega}$,
and $\boldsymbol{\varphi}$ is
$\smash{Q^{\mathrm{st}}(t,\boldsymbol{\epsilon},\boldsymbol{\Omega},\boldsymbol{\varphi})\equiv
Q[\mathscr{S}^{\mathrm{st}}(t,\boldsymbol{\epsilon},\boldsymbol{\Omega},\boldsymbol{\varphi})]}$.
By applying Eq.~\eqref{sym1} twice, it follows that $\smash{
Q^{\mathrm{st}}(t,\boldsymbol{\epsilon},\boldsymbol{\Omega},\boldsymbol{\varphi})}$
is $2\pi$-periodic in all the components of the vector
$\boldsymbol{\varphi}$, i.e.,
$\smash{Q^{\mathrm{st}}(t,\boldsymbol{\epsilon},\boldsymbol{\Omega},\boldsymbol{\varphi}+2\pi
\boldsymbol{u}^{(j)})=Q^{\mathrm{st}}(t,\boldsymbol{\epsilon},\boldsymbol{\Omega},\boldsymbol{\varphi})}$
for $j=1,\dots,N$. In addition, taking in Eq.~\eqref{sym2} $\tau=-t$, one
obtains that
$\smash{Q^{\mathrm{st}}(t,\boldsymbol{\epsilon},\boldsymbol{\Omega},\boldsymbol{\varphi})=
Q^{\mathrm{st}}(0,\boldsymbol{\epsilon},\boldsymbol{\Omega},\boldsymbol{\varphi}+
\boldsymbol{\Omega}t)}$, i.e., the time evolution of
$\smash{Q^{\mathrm{st}}(t,\boldsymbol{\epsilon},\boldsymbol{\Omega},\boldsymbol{\varphi})}$
admits a description in terms of a time-dependent phase vector of the form
$\boldsymbol{\varphi}+ \boldsymbol{\Omega} t$. Taking into account these
two properties and performing a Fourier expansion in
$\boldsymbol{\varphi}$, it is easy to see that the time average of
$Q^\text{st}$ from $0$ to $T$ reads
\begin{equation}
\overline{Q}_{T}(\boldsymbol{\epsilon},\boldsymbol{\Omega},\boldsymbol{\varphi})=\sum_{\boldsymbol{k}\in \mathbb{Z}^N}q_{\boldsymbol{k}}(\boldsymbol{\epsilon},\boldsymbol{\Omega})e^{i\boldsymbol{k}\cdot(
	\boldsymbol{\varphi}+\boldsymbol{\Omega}T/2)}\mathrm{sinc}\left(\frac{\boldsymbol{k}\cdot\boldsymbol{\Omega}T}{2}\right),
	\label{timeaverage}	
\end{equation}
where $\mathrm{sinc}(x)\equiv\sin(x)/x$ denotes the unnormalized sinus
cardinalis, while the centered dot denotes the usual scalar product of
$N$-dimensional vectors, and
\begin{equation}
	q_{\boldsymbol{k}}(\boldsymbol{\epsilon},\boldsymbol{\Omega})=\int_{-\pi}^{\pi}\dots\int_{-\pi}^{\pi}e^{-i\boldsymbol{k}\cdot
			\boldsymbol{\varphi}}Q^{\mathrm{st}}(0,\boldsymbol{\epsilon},\boldsymbol{\Omega},\boldsymbol{\varphi})\prod_{j=1}^N\frac{d\varphi_j}{2\pi}.
	\label{Fouriercoefficients}
\end{equation}
Note that, according to Eqs.~\eqref{sym1} and \eqref{sym2}, the Fourier coefficients
$q_{\boldsymbol{k}}$ satisfy the symmetry property
$q_{\boldsymbol{k}}(\boldsymbol{\epsilon}^{(j)},\boldsymbol{\Omega})=(-1)^{k_j}q_{\boldsymbol{k}}(\boldsymbol{\epsilon},\boldsymbol{\Omega})$
and, hence, can be written in the form
\begin{equation}
\smash{q_{\boldsymbol{k}}(\boldsymbol{\epsilon},\boldsymbol{\Omega})\equiv
\gamma_{\boldsymbol{k}}(\boldsymbol{\epsilon},\boldsymbol{\Omega})\prod_{j=1}^N\epsilon_j^{|k_j|}},
\label{ampdep}
\end{equation}
where the functions
$\smash{\gamma_{\boldsymbol{k}}(\boldsymbol{\epsilon},\boldsymbol{\Omega})}$
are even in each of the arguments $\epsilon_j$. Under quite general
conditions, it can be shown that the functions
$\smash{\gamma_{\boldsymbol{k}}(\boldsymbol{\epsilon},\boldsymbol{\Omega})}$
admit a Taylor expansion in the amplitudes $\epsilon_j$
\cite{CasadoPascualPRE15}.  In practice, for sufficiently small values of
the amplitudes $\epsilon_j$ expressed in suitable dimensionless units, this
expansion can be truncated after a few terms \cite{CasadoPascualPRE15}. The
dependence of the Fourier coefficients on the amplitudes in
Eq.~\eqref{ampdep} can also be obtained by a functional expansion on the
driving~\cite{QuinteroPRE10,CuestaPRX13}.

In the limit $T\to \infty$, the sinc functions appearing in
Eq.~\eqref{timeaverage} vanish unless the resonance condition $\bm
k\cdot\bm\Omega = 0$ is fulfilled. Therefore, Eq.~\eqref{timeaverage} leads to
\begin{equation}
	\overline{Q}_{\infty}(\boldsymbol{\epsilon},\boldsymbol{\Omega},\boldsymbol{\varphi})=\sum_{\boldsymbol{k}\in S_{\bm\Omega}^{\perp}}q_{\boldsymbol{k}}(\boldsymbol{\epsilon},\boldsymbol{\Omega})e^{i\boldsymbol{k}\cdot
		\boldsymbol{\varphi}},
	\label{inf_limit}
\end{equation}
where $S_{\bm\Omega}^{\perp}$ is the set of vectors $\bm k$ that have integer
components and are orthogonal to $\bm \Omega$. In practice, the
limit $T\to\infty$ can be calculated only approximately by taking a
sufficiently large value of $T$.  If we consider the vicinity of a
resonance at a fixed frequency vector $\boldsymbol{\Omega_0}$ and focus on
driving frequencies $\bm \Omega = \bm\Omega_0 +\bm{\delta\omega}$, with
$\bm{\delta\bm\omega}$ of the same order of magnitude as $T^{-1}$,  then
the asymptotic behavior of Eq.~\eqref{timeaverage} for $T\to \infty$ is
given by~\cite{OliveraAtencioEPJB20} 
\begin{equation}
	\begin{split}
		\overline{Q}_{T}(\boldsymbol{\epsilon},\boldsymbol{\Omega_0}+\bm{\delta \omega},\boldsymbol{\varphi}) \sim&\sum_{\boldsymbol{k}\in S_{\bm\Omega_0}^{\perp}}q_{\boldsymbol{k}}(\boldsymbol{\epsilon},\boldsymbol{\Omega_0})
		e^{i\boldsymbol{k}\cdot(
			\boldsymbol{\varphi}+\boldsymbol{\delta \omega}T/2)}\\
		& \times\mathrm{sinc}\left(\frac{\boldsymbol{k}\cdot\boldsymbol{\delta \omega}T}{2}\right).
	\end{split}
\label{asymptotic}
\end{equation}
Rather importantly, owing to the orthogonality condition $\bm k\cdot
\bm\Omega_0 = 0$, the only dependence on $\bm\Omega_0$ is contained in the
Fourier coefficients $q_{\boldsymbol{k}}$.

A non-trivial solution of the equation $\bm k\cdot\bm\Omega = 0$ requires
that the $N$ components of $\bm \Omega$ are commensurable, i.e., that one
of the frequencies can be expressed as a linear combination of the others
with rational coefficients. Otherwise, the set  $S_{\bm\Omega}^{\perp}$
reduces to the trivial solution $\bm{k} =\bm{0}$, and
$q_{\boldsymbol{0}}(\boldsymbol{\epsilon},\boldsymbol{\Omega})$ is the only
non-vanishing term in
$\overline{Q}_{\infty}(\boldsymbol{\epsilon},\boldsymbol{\Omega},\boldsymbol{\varphi})$. Since $q_{\boldsymbol{0}}(\boldsymbol{\epsilon},\boldsymbol{\Omega})$  is independent of the phases in $\bm\varphi$, it provides a smooth
background for sinc-shaped peaks in
$\overline{Q}_{T}(\boldsymbol{\epsilon},\bm\Omega,\bm\varphi)$.
%
%From Eq.~\eqref{Fouriercoefficients} one can easily see that $q_{\bm0}$ is
%independent of the phases in $\bm\varphi$.  Hence, it provides a smooth
%background for sinc-shaped peaks in
%$\overline{Q}_{T}(\boldsymbol{\epsilon},\bm\Omega,\bm\varphi)$.
Interestingly, this background vanishes if the dynamical equations are
invariant under a mapping that involves the phases $\bm\varphi$ and inverts
the sign of the observable $Q$.  Then,
$\overline{Q}_{T}(\boldsymbol{\epsilon},\boldsymbol{\Omega},\boldsymbol{\varphi})
=
-\overline{Q}_{T}(\boldsymbol{\epsilon},\boldsymbol{\Omega},\tilde{\boldsymbol{\varphi}})$,
which links the response for the phase $\bm\varphi$ and the transformed
phase $\tilde{\bm\varphi}$.  If, in addition, the Jacobian of the phase
transformation is unity (which is fulfilled for any phase inversion and
phase shift), one can conclude from Eq.~\eqref{Fouriercoefficients} that
$q_{\bm 0}(\bm\epsilon,\bm\Omega) = -q_{\bm 0}(\bm\epsilon,\bm\Omega) = 0$.

\section{Equivalent phases}

Equation~\eqref{inf_limit} contains a phase factor
$e^{i\bm{k}\cdot\bm{\varphi}}$ which is invariant under a phase shift
$\bm{\varphi} \to \bm{\varphi}+\delta\bm{\varphi}$ provided that
\begin{equation}
\bm{k}\cdot \delta\bm{\varphi} = 2\pi
\label{phases}
\end{equation}
(or any other multiple of $2\pi$) for all the vectors $\bm k$
orthogonal to $\bm \Omega$ and with integer components. In the case that the $N$ components of $\bm \Omega$ are pairwise
commensurable (i.e., 
if there exist a frequency $\Omega_0$ and an $N$-dimensional vector $\bm{n}$ with positive integer components  such that  $\bm{\Omega} =\Omega_0 \bm{n}$), the orthogonality condition $\bm k\cdot \bm\Omega = 0$ becomes equivalent to the Diophantine equation $\bm k\cdot \bm n = 0$. This Diophantine equation together with Eq.~\eqref{phases}  implies invariance of the response under non-trivial phase shifts $\delta\varphi_j<2\pi$.

While for multi-chromatic driving, the general solution of the Diophantine equation $\bm k\cdot \bm n = 0$
may be complicated (see, e.g., Refs.~\cite{MoritoFQ79,MoritoAI80,CasadoPascualPRE15, OliveraAtencioEPJB20}), for bichromatic driving, it can be
derived explicitly. Setting $\bm{\Omega} = (q,p)\Omega_0$ with $p$ and $q$
coprime, the general solution is $(p,-q) \ell$, with $\ell$ being any
integer. Therefore, only terms with $\bm{k}$ being integer multiples of
$(p,-q)$ contribute to $\overline{Q}_\infty$ as discussed above.  Then,
condition~\eqref{phases} becomes $p\delta\varphi_1 - q\delta\varphi_2 = 2\pi$.
Since only relative phases of the drivings $f_j(t)$ matter, we can set one
phase to zero, such that we can conclude invariance of the response
$\overline{Q}_\infty$ for the phase shifts \cite{CasadoPascualPRE13,
OliveraAtencioEPJB20, KohlerEPJB20}
\begin{align}
\varphi_1 \to{}& \varphi_1 + 2\pi/p , \\
\varphi_2 \to{}& \varphi_2 + 2\pi/q .
\end{align}
Notice that this is a kind of cross relation, since the integer $q$ or $p$ that defines the frequency of one of the drivings appears in the phase invariance of the other driving.  In
Ref.~\cite{OliveraAtencioEPJB20}, it has been demonstrated numerically that
the equivalence of these phases holds (approximately) in a whole vicinity
of a $(q,p)$-resonance.

\section{Examples for the bichromatic case}

To illustrate the features discussed so far, we provide explicit numerical
results for a classical random walk and a dissipative quantum mechanical
two-level system. 

\subsection{Classical system} 

Several classical models have been considered in the literature to analyze
generic properties of dissipative dynamical systems under multi-frequency
drivings.  For example,  a one-dimensional model consisting of a Brownian
particle, moving in a periodic potential, under the influence of a
biharmonic force has been used to study some general asymptotic properties
of driven nonlinear dissipative systems  in the long-time
limit~\cite{CasadoPascualPRE13}. This same model has also been considered
to elucidate the connection between irrationality and quasiperiodicity in
these kinds of systems~\cite{CuberoPRL14}. The generality of these results
has been revealed by replacing the periodic potential with a double-well
potential (see the supplemental material in Ref.~\cite{CuberoPRL14}).  In
Ref.~\cite{OliveraAtencioEPJB20}, the generic shape of the resonance peaks
in the vicinity of commensurable frequencies has been illustrated using a
classical random walk model.  Given the simplicity of this latter model, it
will be the one considered here.

As an example for a classical stochastic process, we thus employ an infinite
one-dimensional chain with thermal nearest-neighbor hopping with the
forward and backward rates
\begin{equation}
r_\pm(t) = r_0 e^{-\beta [E_0 \pm \Delta E(t)]},
\end{equation}
where $\beta$ denotes the inverse thermal energy
$1/k_BT$~\cite{OliveraAtencioEPJB20}.  The energy
difference between two adjacent sites with distance $a$ contains a static
contribution $E_0$ and a time-dependent one, $\Delta E(t) = f(t)/\beta$
with
\begin{equation}
f(t) = A_1\cos(\Omega_1 t+\varphi_1) + A_1\cos(\Omega_2 t+\varphi_2) .
\end{equation}
It can be shown that the stationary state of the corresponding master
equation reads
\begin{equation}
v(t) = a[r_+(t) - r_-(t)] = v_0 \sinh[f(t)]
\end{equation}
with $v_0 = 2ar_0 \exp(-\beta E_0)$. For details of the calculation, see
Ref.~\cite{OliveraAtencioEPJB20}.

%--------------
\begin{figure}
\centerline{\includegraphics[width=1.05\columnwidth]{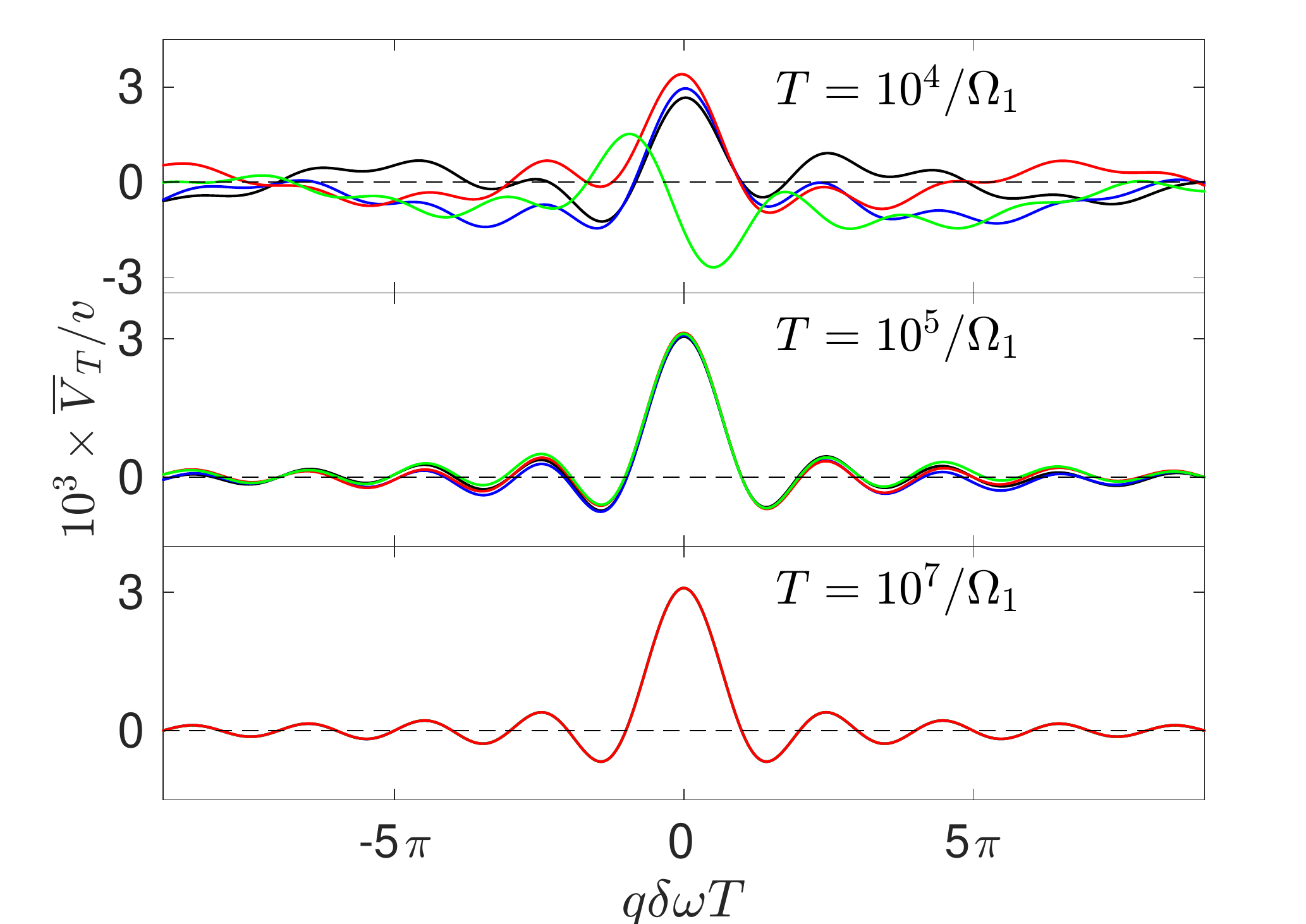}}
\caption{Emergence of the $(q,p) = (4,1)$ resonance peak with increasing averaging
time $T$ for $\varphi_1=0$ and the equivalent phases $\varphi_2 = 0$,
$\pi/2$, $\pi$, $3\pi/2$.  As expected from the theoretical analysis, with
increasing averaging time, all curves converge to the asymptotic behavior in
Eq.~\eqref{asymptotic}.}
\label{fig:buildup}
\end{figure}
%--------------

To evaluate the long-time average of the velocity, it is convenient to
decompose $v(t)$ into a Taylor series in the amplitudes $A_i$.  Then the
time integration of each term can be evaluated analytically, while for the
summation of the resulting terms, we resort to numerics
\cite{OliveraAtencioEPJB20}.  The result for $\Omega_2$ in the vicinity of
$\Omega_1/4$, i.e., close to the (4,1) resonance, is depicted in
Fig.~\ref{fig:buildup}.  It nicely shows that while the response for
equivalent phases defined in the previous section may be different for
small averaging times, all the curves become indistinguishable for
sufficiently large $T$.  Moreover, once convergence is practically reached,
the curves exhibit the sinc shape proposed for their enveloping function.

\subsection{Dissipative two-level system}

In Ref.~\cite{OliveraAtencioEPJB20}, the shape of the resonance peaks
has been investigated also for the two-level Hamiltonian as an example for
dissipative quantum systems.  Here, we consider a quantum mechanical system
defined by the Hamiltonian
\begin{equation}
\begin{split}
H(t) ={}& \frac{\Delta}{2}(\sigma_z\cos\theta + \sigma_x\sin\theta)
\\ & + A_1\sigma_x\cos(\Omega_1 t) + A_2\sigma_z\cos(\Omega_2 t+\varphi)
\end{split}
\label{tls}
\end{equation}
with the Pauli matrices $\sigma_{x,z}$ and the angle $\theta$ which allows the
control of the symmetry, as we will see below.
Dissipation is provided by a Lindblad form such that the quantum master
equation for the density operator reads
\begin{equation}
\dot\rho = -\frac{i}{\hbar}[H(t),\rho]
+\Gamma(2 \sigma_\downarrow\rho\sigma_\uparrow -
\sigma_\uparrow\sigma_\downarrow\rho -\rho \sigma_\uparrow\sigma_\downarrow),\label{LindbladEq}
\end{equation}
with the dissipation rate $\Gamma$ and the Lindblad operator
$\sigma_\downarrow \equiv
|\phi_0\rangle\langle\phi_1| = \sigma_\uparrow^\dagger$, which is the projector to
the ground state $|\phi_0\rangle$ for given angle $\theta$.

%-----------------
\begin{figure}
\centerline{\includegraphics{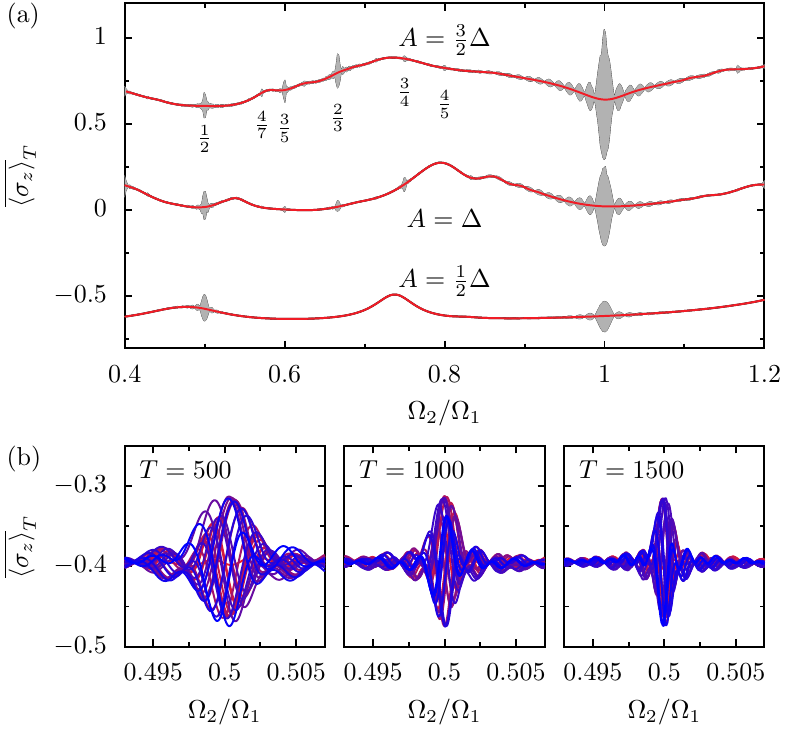}}
\caption{(a) Expectation value $\overline{\langle\sigma_z\rangle}_T$ averaged
over a time $T=1000/\Omega_1$ for all possible phases $\varphi_2$ and
various driving amplitudes (gray) for $\theta=\pi/4$ and
damping rate $\Gamma = 0.05$.  The red lines mark the result of the
two-color Floquet theory with MCF which corresponds to the phase average.
For graphical reasons, the curves for $A=\Delta$ and $A=1.5\Delta$ are
vertically shifted by $0.5$ and $1$, respectively.
(b) Enlargement of the resonance at $\Omega_2=\Omega_1/2$ for $A=1.5\Delta$
and averaging times $T= 500$, $1000$, $1500$ for 25 randomly chosen
relative phases, which visualizes the buildup of the sinc-shaped peak with
increasing $T$.}
\label{fig:tls}
\end{figure}
%-----------------

Figure~\ref{fig:tls}(a) depicts the behavior of the long-time solution for
which we predicted the result in Eq.~\eqref{timeaverage}.
The grey lines show the long-time average
$\overline{\langle\sigma_z\rangle}_T$ for various initial phases.  They are
computed via straightforward numerical propagation of the Lindblad master
equation~\eqref{LindbladEq}.  The curves possess resonance peaks at rational
values of $\Omega_1/\Omega_2$.  Their enveloping function clearly exhibits
the shape of a sinc function, which implies that the series in
Eq.~\eqref{timeaverage} is governed by a single coefficient with index ${\bm
k}\neq{\bm 0}$. The smooth background corresponds to the only coefficient
for which the sinc becomes equal to unity, namely $q_{\bm 0}({\bm\Omega})$.
Figure~\ref{fig:tls}(b) visualizes how the shape of the enveloping
function of the response for different phases emerges.  For
sufficiently large $T$, firstly the sinc-shape is assumed.  Then with $T$
increasing further, the sinc becomes ever narrower and eventually shrinks
to a single discontinuity located at $\Omega_2 = (p/q)\Omega_1$.

As a consequence of spatio-temporal symmetries, the background may vanish.
For example, when $\theta=\pi/2$ the Hamiltonian $H(t)$ is
invariant under unitary transformation with $\sigma_x$ accompanied by a
phase shift $\varphi\to\varphi+\pi$, while our observable $\sigma_z$
acquires a minus sign.  Moreover, since the Lindblad dissipator is defined via
the eigenstates of the time-independent part of the Hamiltonian, it
is invariant under this transformation as well.  Since for incommensurable
frequencies, the phase is irrelevant in the limit $T\to\infty$, the
time-averaged response is equal to its negative value and, hence, must be
zero.

\section{Computation of the phase-averaged response}

An established technique for treating periodically driven systems is Floquet theory.
It is based on the discrete time translation by the period of the driving
$T$. Under this symmetry, linear differential equations $\dot\psi =
L(t)\psi$ possess a complete set of solutions of the form $\psi(t) =
e^{-i\mu t}\phi(t)$, where $\phi(t) = \phi(t+T)$ shares the time
periodicity of the driving \cite{FloquetAENS83}. Then the Floquet function
$\phi(t)$ is an eigen solution of $L(t)-\partial_t$ in a Hilbert space
extended by a periodic time coordinate \cite{ShirleyPR65, SambePRA73}.

For bichromatic driving with incommensurable frequencies, the discrete time
translation symmetry gets lost. Nevertheless, one can employ a Floquet
ansatz extended by a further Fourier index that reflects the periodicity of
the second driving.  Hence the solutions are still of the form $\psi(t) =
e^{-i\mu t}\phi(t)$, but now with the modified Floquet function
\cite{Hanggi98, ChuPR04}
\begin{equation}
\phi(t) = \sum_{\bm k} e^{-i{\bm k}\cdot{\bm\Omega}t} \phi_{\bm
k}({\bm\Omega}) ,
\label{floquet_bichrom}
\end{equation}
where ${\bm k} = (k_1,k_2)$.
It corresponds to a two-dimensional Fourier ansatz which for very strong
driving may be numerically expensive.  Notice that in contrast to the
ansatz for the long-time solution \eqref{timeaverage}, the general solution of the
dynamical system, $\psi(t)$, contains an exponential prefactor.

For dissipative equations of motion, an efficient numerical method has been
developed in Ref.~\cite{ForsterPRB15b}.  It starts from the observation
that the Floquet index $\mu$ of the long-time solution must vanish.  Then
one readily obtains a set of coupled homogeneous equations for the Fourier
coefficients $\phi_{\bm k}({\bm\Omega})$.  The
idea is now to derive for one Fourier index, say $k_2$,
a recurrence equation which can be solved with matrix continued-fraction
(MCF) with a numerical effort that grows only linearly with the size of
the cutoff index.  Finally, this provides the Fourier coefficient
$\phi_{\bm 0}(\bm\Omega)$ which contains all information about the
time-averaged long-time solution.  Hence, it is equivalent to the component
$q_{\bm 0}(\bm\Omega)$ of the long-time solution \eqref{timeaverage}.

While the ansatz \eqref{floquet_bichrom} looks rather natural, it has to be
handled with care, because for commensurable frequencies, it is
overcomplete.  Technically this may lead to divergences in the
MCF iteration. In practice, dissipation generally
cures this problem, but its emergence cannot be ruled out.

Hence, for commensurable frequencies, the MCF algorithm provides the
phase-average of the long-time average \eqref{timeaverage}. In
Fig.~\ref{fig:tls}, we verify this numerically for the case of the
two-level system defined in Eq.~\eqref{tls}. The red curve is computed with
the MCF iteration and indeed provides the smooth background of the
resonance peaks.  A further test may be performed with the mixing angle
$\theta=\pi/2$ for which the symmetry considerations above predict
$\overline{\langle\sigma_z\rangle_T} = 0$.

\section{Proposal for an implementation with trapped ions}

The dissipative dynamics of Eq.~\eqref{LindbladEq} can be carried out
straightforwardly with a trapped ion quantum platform, as has been
demonstrated both
theoretically~\cite{MullerNJP11} and experimentally~\cite{BarreiroN11}.
We consider a two-ion system, where the first ion will encode
the two-level system under study, and the second ion will be an ancillary
qubit that will provide the dissipative part. The unitary part given by
$H(t)$ amounts to a single-qubit time dependent operation, which can always
be decomposed onto single-qubit drivings with appropriate laser intensities
and frequencies~\cite{LeibfriedRMP03}. With respect to the dissipative,
Lindblad-form term of Eq.~\eqref{LindbladEq}, one can couple the previous
two-level system of the first ion with a second two-level system of the
ancillary ion, and perform a digital decomposition of the Lindblad
dynamics, as described in Ref.~\cite{MullerNJP11}. In each digital step, one
would implement the Kraus operators, of the form (considering only the
dissipative part for simplicity, while the unitary part would be carried
out with a subsequent digital step)
\begin{equation} 
\rho(t)=E_0\rho(0)E_0^\dag + E_1\rho(0)E_1^\dag,
\end{equation}
where
\begin{equation}
E_0= \begin{pmatrix} \sqrt{1-\gamma'} & 0 \\ 0 & 1 \end{pmatrix},\quad
E_1= \begin{pmatrix}0 & 0 \\ \sqrt{\gamma'} & 0 \end{pmatrix}
\end{equation}
are 2$\times$2 matrices acting on the two-level subspace of the considered
system with
exponentially decaying $1-\gamma'=\exp(-2 \Gamma t)$. These Kraus operators
correspond to the dissipative channel in the basis $|\phi_0\rangle$,
$|\phi_1\rangle$, providing, in each small time step, the Lindblad part of
the dynamics of Eq.~\eqref{LindbladEq}~\cite{NielsenChuang2000}. To carry out
these operations in a digital quantum simulator with the two-ion system,
for each digital step, one would initialize the ancilla qubit in state
$|0\rangle$ and apply a two-qubit gate $U_2$ such that $\langle
0|U_2|0\rangle=E_0$ and $\langle 1|U_2|0\rangle=E_1$ are the required
matrix elements in the ancillary qubit basis (corresponding to single-qubit
gates in the system qubit). A $U_2$ fulfilling these requirements can
always be obtained via at most three CNOT gates combined with single-qubit
gates~\cite{NielsenChuang2000}. Subsequently, as described in
Refs.~\cite{MullerNJP11,BarreiroN11}, one would apply optical pumping to
the ancillary qubit, to map it to state $|0\rangle$, providing the entropy
increase that realizes the dissipation. Finally, one would carry out the
unitary part of $H(t)$. The complete master equation dynamics would be
provided by the subsequent iteration of this digital step for $n$ total
steps. The long term solution would be obtained for sufficiently large $n$,
and the measurement $\overline{\langle\sigma_z\rangle}_T$ can be
straightforwardly carried out with the trapped ion system via resonance
fluorescence~\cite{LeibfriedRMP03}.

\section{Conclusions and future perspectives}

In this article we have reviewed the generic behavior of
multi-chromatically driven dissipative systems in the classical as well as
in the quantum mechanical dynamics.  Most prominently, in the time-averaged
signal as a function of one driving frequency, one observes phase-dependent
peaks with a sinc-shaped envelope on top of a smooth phase-independent
background.  The width of the peaks diminishes with the averaging time.  By
contrast, the background does not depend on the phases and converges rather
rapidly to its asymptotic value.  Symmetries may suppress the background,
while features of the peaks remain. It is worth mentioning that the width
of these peaks is generally smaller than the one predicted by the Fourier
inequality~\cite{SzriftgiserPRL2002}, where the difference can be expressed as a
factor determined by the driving frequencies~\cite{CasadoPascualPRE13}.
Therefore, our results may have a direct and practical application for the
identification of dissipative systems displaying sub-Fourier
resonances~\cite{CasadoPascualPRE13,CuberoPRL14,SzriftgiserPRL2002}.

To observe these peaks, one has to leave the linear response limit and
enter the regime of harmonic mixing.  Then with an increasing amplitude an
increasing number of resonances emerges, but ``simple resonances'' such as
1/1, 1/2, or 2/3 dominate the overall behavior.  A challenging open problem
is the question whether there is any rule for the relative magnitude of the
peaks as a function of the ``simplicity'' of the frequency ratio.

For the computation of the numerical examples, we have used simple
propagation schemes which, however, may be rather time consuming.  For the
phase-averaged response, it turned out that a two-frequency Floquet theory
provides reliable results for practically all frequency ratios, despite
that its convergence is guaranteed only for incommensurable frequencies.
This allows one to employ a computational method based on matrix
continued-fractions, which is numerically rather efficient.

Finally, future explorations of the magnitudes of the resonances for
further systems may deepen our understanding of multi-chromatically
driven systems and may open perspectives in science, technology, and
industry.

\acknowledgments
We acknowledge funding by the Junta de Andaluc\'ia (P20-00617 and  US-1380840) and by the
Spanish Ministry of Science, Innovation, and Universities under grant Nos.\ FIS2017-86478-P,
PGC2018-095113-B-I00, PID2019-104002GB-C21, PID2019-104002GB-C22,
PID2020-117787GB-I00 (MCIU/AEI/FEDER, UE), and via the CSIC Research
Platform on Quantum Technologies PTI-001.

\bibliographystyle{eplbib}

\end{document}